\documentclass[12pt]{article}
\pdfoutput=1

\usepackage{amsmath} 
\usepackage{amssymb} 
\usepackage[font={small}]{caption} 
\usepackage{color} 
\usepackage{hyperref} 
\usepackage[letterpaper,margin=1in,bottom=1in]{geometry} 
\usepackage{graphicx} 
\usepackage{subcaption} 
\usepackage[numbers, sort & compress]{natbib} 
\usepackage{parskip} 
\usepackage[section]{placeins} 

\author{
  Pran Nath\footnote{Email: p.nath@northeastern.edu}~\ and
  Maksim Piskunov\footnote{Email: m.piskunov@northeastern.edu}\\~\\
  Department of Physics, Northeastern University,\\
  Boston, MA 02115-5000, USA
}

\title{
  Enhancement of the Axion Decay Constant in Inflation and the Weak Gravity Conjecture
}

\begin{document}
\maketitle
\date

\textbf{Abstract:}
Models of axion inflation based on a single cosine potential require the axion decay constant $f$ to be super-Planckian in size.
However, $f > M_{Pl}$ is disfavored by the Weak Gravity Conjecture (WGC).
It is then pertinent to ask if one can construct axion inflation models in conformity with WGC.
In this work we assume that WGC holds for the microscopic Lagrangian so that $f < M_{Pl}$.
However, inflation is controlled by an effective Lagrangian much below the Planck scale where the inflaton is an effective axionic field associated with an effective decay constant $f_e$ which could be very different from $f$.
In this work we propose a Coherent Enhancement Mechanism (CEM) for slow roll inflation controlled by flat potentials which can produce $f_e \gg M_{Pl}$ while $f < M_{Pl}$.
In the analysis we consider a landscape of chiral fields charged under a $U\left(1\right)$ global shift symmetry and consider breaking of the $U\left(1\right)$ symmetry by instanton type symmetry breaking terms.
In the broken phase there is one light pseudo-Nambu-Goldstone-Boson (pNGB) which acts as the inflaton.
We show that with an appropriate choice of symmetry breaking terms the inflaton potential is a superposition of many cosines and the condition that they produce a flat potential allows one to enhance $f_e$ so that $f_e / M_{Pl} \gg 1$.
We discuss the utility of this mechanism for a variety of inflaton models originating in supersymmetry and supergravity.
The Coherent Enhancement Mechanism allows one to reduce an inflation model with an arbitrary potential to an effective model of natural inflation, i.e. with a single cosine, by expanding the potential near a field point where horizon exit occurs, and matching the expansion coefficients to those of natural inflation.
We demonstrate that this approach can predict the number of e-foldings in a given inflation model without the need for numerical simulation.
Further we show that the effective decay constant $f_e$ can be directly related to the spectral indices so that $f_e = M_{Pl} / \sqrt{1 - n_s - r / 4}$ where $n_s$ is the spectral index for curvature perturbations and $r$ is the ratio of the power spectrum of tensor perturbations and curvature perturbations.
The current data on $n_s$ and $r$ constrains the effective axion decay constant so that $4.9 \leq f_e / M_{Pl} \leq 10.0$ at $95\%$ CL.
Thus an important result of the analysis is that the effective axion decay constant has an upper limit of $\sim 10 M_{Pl}$ in axion cosmology for any potential-based model which produces successful inflation.
For the Dirac-Born-Infeld inflation and more generally k-flation CEM is not applicable.
Nonetheless in this case also we show that successful inflation can occur with $f < M_{Pl}$.
Further, one can define slow-roll parameters as well as the effective axion decay constant in terms of inflaton density which is valid both for models using potentials as well as DBI-flation and more generally k-flation.
In the models considered in this work, all the moduli are stabilized and the inflation model in each case is consistent with astrophysical observations with $f_e > M_{Pl}$ and the axion decay constant of the microscopic theory $f < M_{Pl}$ consistent with the Weak Gravity Conjecture.
In conclusion, among the models we considered those with flat potentials and consistent with WGC have $r < O\left(10^{-3}\right)$, and the only single field model we considered consistent with WGC and $r$ as large as the experimental limit $r = 0.07$ is the DBI model.
\footnote{Source code: \url{https://github.com/maxitg/coherent-enhancement}}
\newpage

\section{Introduction \label{sec:Introduction}}
As is well known many of the problems associated with Big Bang cosmology which include
the flatness problem, the horizon problem, and the monopole problem are resolved by inflation~\cite{Guth:1980zm, Starobinsky:1980te, Linde:1981mu, Albrecht:1982wi, Linde:1983gd}.
Quantum fluctuations at the time of horizon exit carry significant information regarding specifics of the inflation model~\cite{Mukhanov:1981xt, Hawking:1982cz, Starobinsky:1982ee, Guth:1982ec, Bardeen:1983qw, Cheung:2007st} which can be extracted from cosmic microwave background (CMB) radiation anisotropy.
The data from the Planck experiment~\cite{Akrami:2018vks, Akrami:2018odb, Array:2015xqh} has helped constrain inflation models excluding some and narrowing down the parameter space of others.
One such model is the so-called natural inflation based on a $U(1)$ shift symmetry which is described by a simple potential~\cite{Freese:1990rb, Adams:1992bn} $V\left(a\right) = \Lambda^4 \left(1 - \cos\left(\frac{a}{f}\right)\right)$, where $a$ is the axion field and $f$ is the axion decay constant.
In this case, consistency with Planck data requires the axion decay constant to be significantly greater than the Planck mass $M_{Pl}$.
However, an axion decay constant larger than the Planck mass is undesirable since a global symmetry is not preserved by quantum gravity unless it has a gauge origin.
Additionally string theory prefers the axion decay constant to lie below $M_{Pl}$~\cite{Banks:2003sx, Svrcek:2006yi}.
These results are codified in WGC~\cite{ArkaniHamed:2006dz} which requires $f < M_{Pl}$.
It is then relevant to ask if in general axion inflation models can be constructed consistent with the WGC constraint.

In this work we show that one can in fact construct axionic inflation models consistent with the WGC constraint and consistent with data within supersymmetry and supergravity models~\cite{Nath:2017ihp} and supersymmetric Dirac-Born-Infeld models~\cite{Nath:2018xxe} (for a review of supersymmetric inflation see, e.g.,~\cite{Nath:2016qzm}).
Before going into details it is to be noted that the WGC constraint applies to the microscopic theory where axion is a primary field.
On the other hand inflation is driven by an effective theory far below the Planck scale, where the inflaton is an effective axion field in the effective theory which exists in some broken symmetry phase.
Such a situation occurs if one considers a landscape of chiral fields each of which are charged under a global $U\left(1\right)$ symmetry which is broken by instantons.
In this case the inflaton is an effective axion field which is a linear combination of many axion fields and is associated with an effective axion decay constant which can be very different from the axion decay constant of the microscopic theory.
Several suggestions along this line exist such as the alignment mechanism~\cite{Kim:2004rp, Long:2014dta}.
There is a significant amount of literature associated with this topic, see, e.g.,~\cite{Rudelius:2015xta, Rudelius:2014wla, Bachlechner:2014gfa, Choi:2014rja, delaFuente:2014aca, Blumenhagen:2014gta, Hebecker:2015rya, Conlon:2016aea, Montero:2015ofa, Junghans:2015hba}.
Here we propose a new mechanism, i.e., the Coherent Enhancement Mechanism (CEM), where slow roll is governed by a flat potential which allows the effective decay constant $f_e \gg M_{Pl}$ while the primary decay constant $f < M_{Pl}$ consistent with WGC~\cite{Nath:2017ihp}.
The proposed mechanism applies to supersymmetric and supergravity theories.
For Dirac-Born-Infeld-infation CEM does not work.
However, it is shown that DBI can produce successful inflation with $f < M_{Pl}$.
Thus an analysis within these models shows that one can obtain spectral indices as well as the ratio of the tensor to the scalar power spectrum consistent with the Planck data~\cite{Akrami:2018vks, Akrami:2018odb, Array:2015xqh} and consistent with WGC.

The outline of the rest of the paper is as follows.
In section \ref{sec:WeakGravityConjecture} we give a brief discussion of the Weak Gravity Conjecture.
In section \ref{sec:CoherentEnhancement} we discuss the Coherent Enhancement Mechanism when the potential consists of a superposition of cosines which is typically the case for axionic potentials.
In section \ref{sec:Supersymmetry} we discuss inflation for globally supersymmetric models, and in section \ref{sec:Supergravity} for supergravity models and their consistency with WGC.
In section \ref{sec:DBI}, we discuss the Dirac-Born-Infeld case, in which inflation is not controlled by potential alone but by the full Lagrangian.
However, an effective decay constant $f_{eH}$ can still be defined based on inflation dynamics.
In section \ref{sec:r}, we discuss the reason why under the WGC constraint the ratio $r$ of tensor-to-scalar power spectra is $O\left(10^{-4}\right)$ for single field inflation for models with a flat potential while it can be much larger in DBI-flation up to the current experimental limit of $r = 0.07$.
A simple explanation of this phenomenon is seen when one expresses the slow-roll parameters and the spectral indices in terms of inflaton density.
Conclusions are given in section \ref{sec:Conclusion}.
Some relevant papers related to this work can be found in \cite{BlancoPillado:2006he, Conlon:2005jm, Ben-Dayan:2014lca, Gao:2014uha}.

\section{The Weak Gravity Conjecture and axion inflation \label{sec:WeakGravityConjecture}}
In its simplest form the Weak Gravity Conjecture considers the coupling of an abelian gauge field with gravity and states that this system must contain a particle of charge $q$ and mass $m$ so that~\cite{ArkaniHamed:2006dz} $\frac{q}{m} > \frac{1}{M_{Pl}}$, where $M_{Pl}$ is the reduced Planck mass defined by $M_{Pl} = \left(8 \pi G_N\right)^{-1 / 2}$ and $G_N$ is Newton's constant.
The existence of such a particle is needed to carry away the charge of a black hole to avoid the remnant problem when the black hole evaporates due to Hawking radiation.
The above constraint is found to be consistent with string theory and thus one might argue that consistent theories of quantum gravity obey the Weak Gravity Conjecture.
Specifically, for example, one cannot let the charge $q$ become arbitrarily small since in that case in the limit one will have a continuous global symmetry which is forbidden by strings.
So far the analysis concerns just the abelian gauge theories coupling with gravity.
However, there is a generalized WGC which has implications for axions and for axionic inflation.

The generalized WGC constraints the axion decay constant so that $f \leq M_{Pl} / S$ where $S$ is the instanton action~\cite{Brown:2015iha, Brown:2015lia, Heidenreich:2015wga}.
String theory requires $S \geq 1$, so that the theory is in the perturbative domain, which gives $f \leq M_{Pl}$.
We note in passing that the constraints of WGC for axions are more indirectly arrived at relative to the original WGC.
However, there is support for the generalized conjecture as it relates to axions.
Thus even before WGC, Bank et al.~\cite{Banks:2003sx} analyzed a number of periodic fields in string theory for various string vacua and found that an axion decay constant larger than the Planck mass was undesirable.
Also analyses for a wide variety of axions in strings estimate the axion decay constant to lie in the range $\left(10^{16} - 10^{18}\right) \text{GeV}$~\cite{Svrcek:2006yi}.

WGC poses a problem for natural inflation since natural inflation requires $f > 5 M_{Pl}$ in apparent contradiction with WGC.
However, here we need to keep in mind that the WGC constraint on the axion decay constant applies to the microscopic theory.
An effective theory below the Planck scale is not necessarily subject to that constraint.
More specifically, the inflaton need not be a primary field in the microscopic theory but rather an effective field such as a linear combination of the primary field in the domain where the $U\left(1\right)$ global symmetry is spontaneously broken and the inflaton possesses an effective potential generated solely from such breaking.
In this case the inflaton would be an axion with an effective axion decay constant which could be signicantly different from the primary one.
This is demonstrated in the next section where we discuss the Coherent Enhancement Mechanism.

\section{General analysis of Coherent Enhancement Mechanism \label{sec:CoherentEnhancement}}
As mentioned in section \ref{sec:Introduction}, Coherent Enhancement Mechanism works for slow-roll inflation arising from a flat potential.
Before discussing this mechanism we derive a relation that gives the effective axion decay constant directly in terms of the slow-roll parameters and in terms of the experimentally measurable spectral indices and the tensor to scalar ratio $r$ of the power spectrum.
Thus we consider a Lagrangian of the form
\begin{equation}
  \mathcal{L}\left(\phi, \dot{\phi}\right) = \frac{1}{2}{\dot{\phi}}^2 - V\left(\phi\right)\,.
\end{equation}
where the kinetic term is canonically normalized.
Our focus is on that part of $V\left(\phi\right)$ where the potential is flat leading to inflation.
We are specifically interested at the point of horizon exit $\phi \sim \phi_0$.
Here it is sufficient to have $V\left(\phi\right) \approx V_{e}\left(\phi\right)$ at $\phi \approx \phi_0$ where
\begin{equation} \label{eq:naturalInflationPotential}
  V_e\left(\phi\right) = \Lambda^4 \left(1 - \cos\left(\frac{\phi}{f_e}\right)\right)\,,
\end{equation}
to have similar evolution of the field and the scale factor near $\phi \sim \phi_0$.
Using this observation, we can express the parameters of natural inflation $\Lambda$ and $f_e$ in terms of various order derivatives of $V\left(\phi\right)$.
To do that, we expand $V_{e}$ near $\phi_0$ to the second order so that
\begin{equation} \label{eq:naturalInflationSeries}
  \begin{aligned}
    V_{e}\left(\phi\right) =
      &\Lambda^4 \left(1 - \cos\left(\frac{\phi_0}{f_e}\right)\right)
        + \frac{\Lambda^4}{f_e} \sin\left(\frac{\phi_0}{f_e}\right) \left(\phi - \phi_0\right)\\
      & + \frac{\Lambda^4}{2 f_e^2} \cos\left(\frac{\phi_0}{f_e}\right) \left(\phi - \phi_0\right)^2
        + \Lambda^4 \mathcal{O}^3\left(\frac{\phi - \phi_0}{f_e}\right)\,.
  \end{aligned}
\end{equation}

Now, identifying the expansion coefficients in Eq.~(\ref{eq:naturalInflationSeries}) with corresponding derivatives of $V\left(\phi\right)$, and solving for $\Lambda$, $f_e$ and $\cos\left(\phi_0 / f_e\right)$, we obtain
\begin{equation} \label{eq:lambdaFromPotential}
  \Lambda^4 = V\left(\phi_0\right) \frac
    {{V^\prime}^2\left(\phi_0\right) - V\left(\phi_0\right) V^{\prime\prime}\left(\phi_0\right)}
    {{V^\prime}^2\left(\phi_0\right) - 2 V\left(\phi_0\right) V^{\prime\prime}\left(\phi_0\right)}
  \,,
\end{equation}
\begin{equation} \label{eq:feFromPotential}
  f_e = \frac
    {V\left(\phi_0\right)}
    {\sqrt{{V^\prime}^2\left(\phi_0\right)
      - 2 V\left(\phi_0\right) V^{\prime\prime}\left(\phi_0\right)}}\,,
\end{equation}
\begin{equation} \label{eq:fieldInitialFromPotential}
  \cos\left(\frac{\varphi_0}{f_e}\right) = \frac
    {V\left(\phi_0\right) V^{\prime\prime}\left(\phi_0\right)}
    {{V^\prime}^2\left(\phi_0\right) - V\left(\phi_0\right) V^{\prime\prime}\left(\phi_0\right)}\,.
\end{equation}

Further, it is convenient to express the effective decay constant $f_e$ in terms of slow-roll inflation parameters $\epsilon$ and $\eta$ defined as
\begin{equation} \label{eq:epsEtaFromPotential}
  \epsilon =
    \frac{M_{Pl}^2}{2}
    \left(\frac{V^\prime\left(\phi_0\right)}{V\left(\phi_0\right)}\right)^2\,,
  ~~~ \eta = M_{Pl}^2 \frac{V^{\prime\prime}\left(\phi_0\right)}{V\left(\phi_0\right)}\,.
\end{equation}

By combining these with Eqs.~(\ref{eq:lambdaFromPotential}, \ref{eq:feFromPotential}, \ref{eq:fieldInitialFromPotential}), we obtain
\begin{align} 
  \label{eq:lambdaSlowRoll}
  \Lambda^4 &= V\left(\phi_0\right) \frac{2 \epsilon - \eta}{2 \epsilon - 2 \eta}\,,\\
  \label{eq:feSlowRoll}
  f_e &= \frac{M_{Pl}}{\sqrt{2 \left(\epsilon - \eta\right)}}\,,\\
  \label{eq:fieldInitialSlowRoll}
  \cos\left(\frac{\varphi_0}{f_e}\right) &= \frac{\eta}{2 \epsilon - \eta}\,.
\end{align}

The spectral indices $n_s$ and $n_t$ are related to the slow-roll parameters so that
\begin{equation} \label{eq:observablesSlowRoll}
  n_s = 1 - 6 \epsilon + 2 \eta\,,
  ~~~ n_t = -2 \epsilon\,,
  ~~~ r = 16 \epsilon\,.
\end{equation}
We can thus eliminate $\eta$ and $\epsilon$ in favor of $n_s$ and $r$ and get
\begin{equation} \label{eq:feSpectralIndices}
  f_e = \frac{M_{Pl}}{\sqrt{1 - n_s - r / 4}}\,.
\end{equation}
The current experimental limits from Planck experiment at $k_0 = 0.05\,{\rm Mpc}^{-1}$ are as follows~\cite{Akrami:2018vks, Akrami:2018odb, Array:2015xqh}
\begin{equation} \label{data}
  \begin{aligned}
    n_s &= 0.9649 \pm 0.0042\, \left(68\%\, {\rm CL}\right)\,,\\
      r &< 0.064\, \left(95\%\, {\rm CL}\right)\,,
  \end{aligned}
\end{equation}
while $n_t\left(k_0\right)$ is currently not constrained.
Using this data we find model-independent bounds on the effective axionic decay constant so that
\begin{equation} \label{eq:feExperimentalConstraint}
  4.9 \leq f_e / M_{Pl} \leq 10.0\, \left(95\%\, {\rm CL}\right)\,.
\end{equation}

Next, we discuss the Coherent Enhancement Mechanism arising from a superposition of cosine functions.
As a specific simple example let us consider a potential of the form
\begin{equation} \label{eq:cosineSumPotential}
  V = \sum_{k = 1}^n \Lambda_k^4 \left(1 - \cos\left(\frac{k\phi}{f}\right)\right)\,.
\end{equation}
Here we choose $\phi_0$ where the maximum occurs so that $\frac{\phi}{f} = \pi$.
On using Eq.~(\ref{eq:feFromPotential}) we get
\begin{equation} \label{eq:feForCosineSum}
  {f_e} / f = \frac
    {\sqrt{\sum_{k \in odd} \Lambda_k^4}}
    {\sqrt{\sum_{k \in odd} k^2 \Lambda_k^4 - \sum_{k \in even} k^2 \Lambda_k^4}}\,.
\end{equation}
One notices that there is a cancellation between the odd and even sums in the denominator in Eq.~(\ref{eq:feForCosineSum}) which leads to an enhancement and gives $f_e / f > 1$.
Since the enhancement occurs as a consequence of the sum of several terms we call this a ``Coherent Enhancement Mechanism''.

It is also interesting to note that $n_s$, $r$, and therefore $f_e$ can be directly computed from the scale factor evolution without invoking any knowledge about the potential.
This formulation is of relevance when we discuss DBI-flation where slow roll is not driven by the potential alone but by the entire Lagrangian.
To distinguish the effective decay constant computed in this way from $f_e$ that is computed from potential, we call it $f_{eH}$.

Specifically, we can define the dynamic slow-roll parameters
\begin{equation} \label{eq:slowRollParametersDynamic}
  \epsilon_H = -\frac{\dot H}{H^2}\,,
  ~~~ \eta_H = \frac{\dot{\epsilon_H}}{\epsilon_H H}\,,
  ~~~ \sigma_H = \frac{1}{H} \frac{d}{dt} \ln c_s\,,
\end{equation}
where $c_s$ is the speed of sound in the medium where $c_s^2 = p_{,\beta} / \rho_{,\beta}$, $\beta = \dot\phi^2$ where $p$ is the pressure and $\rho$ the density.
In this case spectral indices are given by~\cite{Garriga:1999vw, Spalinski:2007qy}
\begin{equation} \label{eq:observablesFromDynamicSlowRoll}
      n_s = 1 - 2 \epsilon_H - \eta_H - \sigma_H,
  ~~~ n_t = - 2\epsilon_H\,.
\end{equation}

For the case when the dependence of the parameters on sound speed $c_s$ is relatively small one has
\begin{equation} \label{eq:slowRollParametersDynamicFromStatic}
  \epsilon_H = \epsilon\,,
  ~~~ \eta_H = -2 \eta + 4 \epsilon\,,
\end{equation}
in which case Eq.~(\ref{eq:feSlowRoll}) can be rewritten as
\begin{equation} \label{eq:feFromDynamicSlowRollParameters}
  f_{eH} = \frac{M_{Pl}}{\sqrt{\eta_H - 2 \epsilon_H}}\,.
\end{equation}

We will show using numerical simulations that $f_e \approx f_{eH}$ for the cases of global supersymmetry and supergravity.
For DBI-flation CEM does not work as there is no analogue of $f_e$ since inflation in not controlled by the potential alone but by the full Lagrangian as mentioned earlier.
However, for DBI one may still define an $f_{eH}$ as given by Eq.~(\ref{eq:feFromDynamicSlowRollParameters}) which may be compared to the true axion decay constant that enters in the DBI lagrangian.
This will be discussed in further detail in section \ref{sec:DBI}.

\section{Global supersymmetry \label{sec:Supersymmetry}}
In this section we consider inflation in globally supersymmetric models where slow roll  is controlled by a flat potential and CEM can operate.
For the analysis here we consider a chiral field $\Phi$ charged under a global $U\left(1\right)$ transformation, and another field $\bar\Phi$ that is oppositely charged.
Thus under $U\left(1\right)$ transformations one has
\begin{equation}
  \Phi \to e^{i q \lambda} \Phi\,,
  ~~~ \bar\Phi \to e^{-i q \lambda} \bar\Phi\,.
\end{equation}
The superfield $\Phi$ has an expansion, $\Phi = \phi + \theta \chi + \theta \theta F$, where $\phi$ is a complex scalar field consisting of the saxion (the magnitude) and the axion (the phase), $\chi$ is the axino, and $F$ is an auxiliary field.
Similarly the superfield $\bar\Phi$ has an expansion: $\bar\Phi = \bar\phi + \bar\theta \bar\xi + \bar\theta \bar\theta \bar F$.
We now consider a superpotential of the form
\begin{equation} \label{eq:supersymmetry:W}
  W = W_s\left(\Phi, \bar\Phi\right) + W_{sb}\left(\Phi, \bar\Phi\right)\,,
\end{equation}
where $W_s$ is the part that depends on the fields $\Phi, \bar\Phi$ and is invariant under the shift symmetry, and $W_{sb}$ is a part which breaks the shift symmetry.
$W_s$ is chosen to stabilize the real parts of the chiral fields and we expand the chiral fields around the stabilized VEVs.
We take $W_s$ of the form
\begin{equation}
  W_s\left(\Phi, \bar\Phi\right) =
    \mu \Phi \bar\Phi + \frac{\lambda}{2 M} \left(\Phi \bar\Phi\right)^2\,.
\end{equation}
We may parametrize $\phi$ and $\bar\phi$ so that
\begin{equation}
  \phi = \left(f + \rho\right) e^{i a / f}\,,
  ~~~ \bar\phi = \left(\bar f + \bar\rho\right) e^{i \bar a / \bar f}\,,
\end{equation}
where $f = \left<\phi\right>$, $\bar f = \left<\bar\phi\right>$ and $\left(\rho, a\right)$ and $\left(\bar\rho, \bar a\right)$ are the fluctuations of the quantum fields around their vacuum expectation values $f$, $\bar f$.
We define two linear combinations of $a$ and $\bar a$ so that
\begin{equation} \label{eq:b+-}
  b_{\pm}= \frac{1}{\sqrt 2} \left(a \pm \bar a\right)\,.
\end{equation}
Here $b_+$ is invariant under the shift symmetry and becomes heavy after the moduli are stabilized and $b_-$ is sensitive to shift symmetry and remains massless and we identify it as a candidate for the inflaton.

However, $b_-$ will gain mass when $W_{sb}$ is included in the analysis.
We take $W_{sb}$ of the form
\begin{equation} \label{eq:supersymmetry:Wsb}
  W_{sb}\left(\Phi, \bar\Phi\right) =
      \sum_{l = 1}^q A_l \Phi^l
    + \sum_{l = 1}^q \bar A_l \bar\Phi^l\,,
\end{equation}
which violates the shift symmetry.
Here we note that a similar procedure of using several non-perturbative terms to produce inflation by adjustment of parameters in the non-perturbative terms is used in the so-called `racetrack' models (see, e.g.,~\cite{BlancoPillado:2004ns, Lalak:2005hr, Greene:2005rn, BlancoPillado:2006he}).
Including $W_{sb}$ the axionic potential can be written in the form
\begin{equation}
  V\left(a, \bar a\right) = V_\text{fast}\left(b_+\right) + V_\text{slow}\left(b_-\right)\,,
\end{equation}
where $V_\text{slow}\left(b_-\right)$ which depends only on $b_-$ enters in slow roll and is relevant for inflation.

\begin{figure}
  \centering
  \begin{subfigure}{0.5 \textwidth}
    \includegraphics[width = \textwidth]{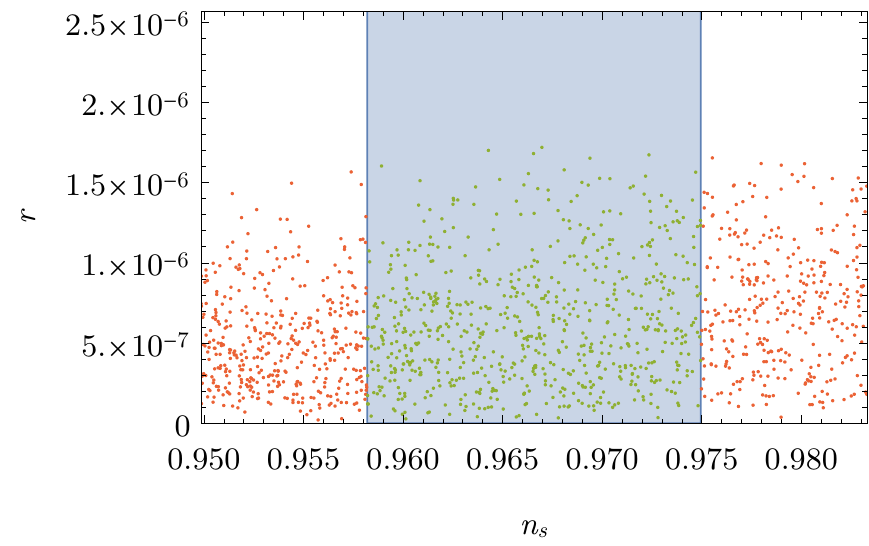}
  \end{subfigure}
  \begin{subfigure}{0.5 \textwidth}
    \includegraphics[width = \textwidth]{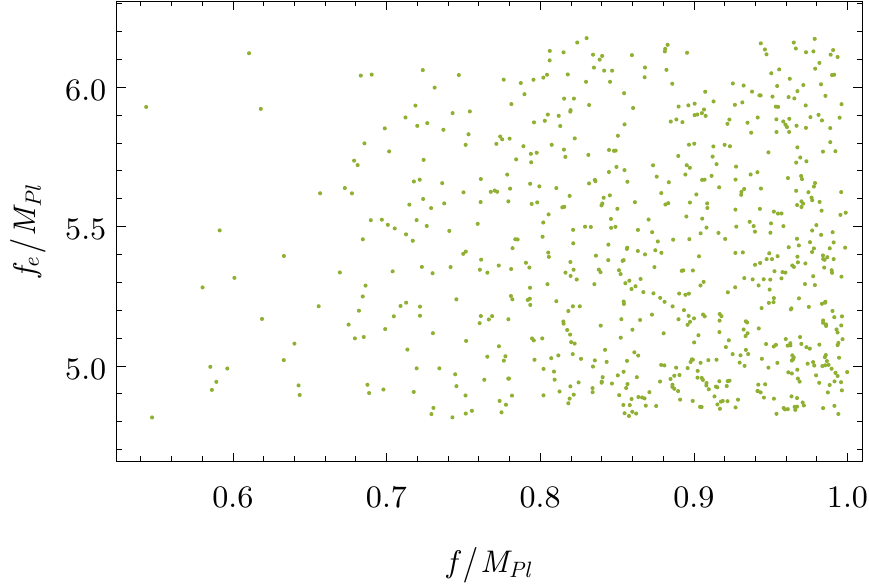}
  \end{subfigure}
  \begin{subfigure}{0.5 \textwidth}
    \includegraphics[width = \textwidth]{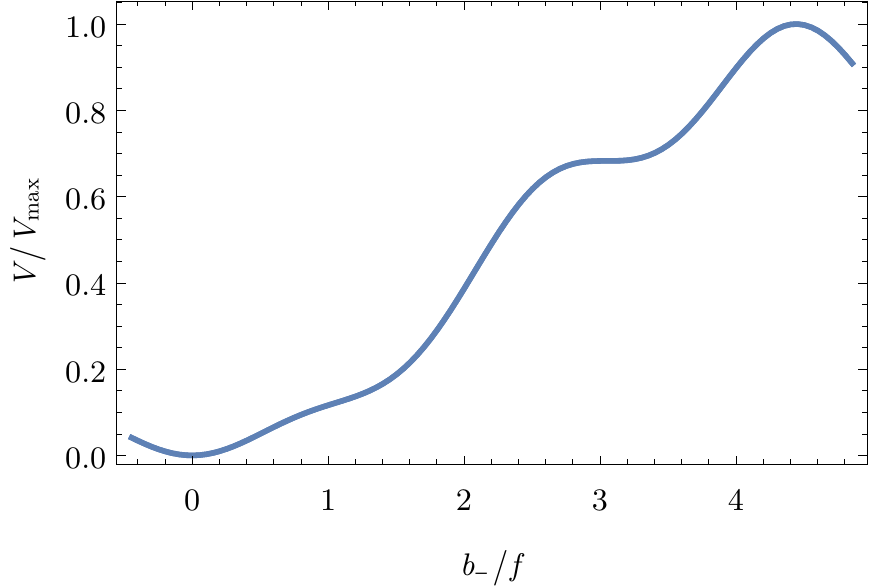}
  \end{subfigure}
  \caption{\protect\input{figures/supersymmetry.txt}
    Top panel shows tensor-to-scalar ratio vs scalar spectral index.
    Blue region encloses the parameter points consistent with Planck 2018 TT,TE,EE+lowE+lensing+BK14+BAO data at $95\%$ CL.
    Middle panel exhibits the coherent enhancement of the decay constant where $f_e \gg M_{Pl}$ while $f < M_{Pl}$.
    The bottom panel shows the superimposed slow-roll potentials~Eq.(\ref{eq:supersymmetry:Vslow}) as functions of $b_-$ for all values of $G_5$ considered.
    Note that because the field is normalized by $f$ and because $G_5$ is fine-tuned, potentials for all considered input parameters look almost identical.
    Inflation occurs in the flat region of the potential near $b_- / f \approx 3$.
    Here the field transversal during inflation is $\Delta b_- < f \le M_{Pl}$.
    The analysis shows that axion inflation for the parameter points in the blue region is consistent with WGC as exhibited by the middle panel.
  } \label{fig:supersymmetry}
\end{figure}

Note, that the parameters $\mu$ and $\lambda / M$ determine the stability point $\left<\phi\right> = f$~\cite{Halverson:2017deq}.
We can equivalently, however, fix $f$ and solve for $\lambda / M$ in terms of $\mu$ and $f$.
It turns out that $\mu$ only appears in $V_\text{fast}$, but not in $V_\text{slow}$, for which an explicit form is given by
\begin{equation} \label{eq:supersymmetry:VslowGeneral}
  \begin{aligned}
    V_\text{slow}\left(b_-\right) =
      &2 \sum_{r = 1}^q r
        \left(A_r f^{r - 1} \sum_{l = 1}^q l A_l f^{l - 1}
          + \bar A_r f^{r - 1} \sum_{l = 1}^q l \bar A_l f^{l - 1}\right)
        \left(1 - \cos\left(\frac{r}{\sqrt 2 f} b_-\right)\right)\\
      &{} - 2 \sum_{l = 1}^q \sum_{r = l + 1}^q
        l r \left(A_l A_r f^{l + r - 2} + \bar A_l \bar A_r f^{l + r - 2}\right)
        \left(1 - \cos\left(\frac{r - l}{\sqrt 2 f} b_-\right)\right)\,,
  \end{aligned}
\end{equation}
where we have set $\bar f = f$.
We make now further simplifying assumptions so that $A_l = \bar A_l = B_l f^{3 - l}$, $B_l = B G_l$.
Thus $B_l$, $B$, $G_l$ are dimensionless while $f$ carries dimension of mass.
Using the above assumptions the potential of Eq.~(\ref{eq:supersymmetry:VslowGeneral}) takes a simpler form
\begin{equation} \label{eq:supersymmetry:Vslow}
  \begin{aligned}
    V_\text{slow}\left(b_-\right) = 4 f^4 B^2 &\left(
      \sum_{l = 1}^q l G_l \sum_{r = 1}^q r G_r
        \left(1 - \cos\left(\frac{r}{\sqrt{2}} \frac{b_-}{f}\right)\right)\right. \\
      &\left.{} - \sum_{l = 1}^q \sum_{r = l + 1}^q l r G_l G_r
        \left(1 - \cos\left(\frac{r - l}{\sqrt{2}} \frac{b_-}{f}\right)\right)
    \right)\,.
  \end{aligned}
\end{equation}

Here the superposition of several cosines produces local flatness where slow roll can occur.
In order to verify consistency with experiment and evaluate $f_e$, we use Inflation Simulator\footnote{\url{https://github.com/maxitg/InflationSimulator}}.
For these simulations we begin by sampling a number of parameter sets as described in the caption of Fig.~(\ref{fig:supersymmetry}).
We then use the Lagrangian $\mathcal{L} = \frac{1}{2} \dot b_-^2 - V_\text{slow}\left(b_-\right)$ and Friedmann equations described in section~4 of~\cite{Nath:2018xxe} to simulate evolution of the field and the scale factor.
We set initial field velocity $\dot b_{-, \text{init}} = 0$, and we only select points where $N_\text{subhorizon} \geq 5$ where $N_\text{subhorizon} = N_\text{total} - N_\text{pivot}$ to ensure that the result of the analysis is not affected by initial conditions.
Finally, if we have sufficient number of e-foldings, we compute the tensor-to-scalar ratio $r$, and the scalar spectral index $n_s$ at horizon exit (i.e., $N_\text{pivot}$ e-foldings before the end of inflation) using Eqs.~(\ref{eq:slowRollParametersDynamic}, \ref{eq:slowRollParametersDynamicFromStatic}, \ref{eq:observablesSlowRoll}), and check if they are consistent with experimental constraints~\cite{Akrami:2018odb}.
If so, we evaluate $f_e$ and $f_{eH}$ at horizon exit using Eqs.~(\ref{eq:feFromPotential}, \ref{eq:feFromDynamicSlowRollParameters}).
The result of this analysis is displayed in Fig.~(\ref{fig:supersymmetry}).
Here one finds that while the true decay constant $f$ is below $M_{Pl}$, the effective $f_e \approx f_{eH}$ always satisfies the constraint \ref{eq:feExperimentalConstraint}.
Further, we find the relative difference between the two effective decay constants $f_e$ and $f_{eH}$ small over the parameter space investigated, i.e., $\left|f_e - f_{eH}\right| / f_e \le \protect 3\%$.
We note that fine tuning of $G_5$ is required to achieve coherent enhancement and experimentally-consistent inflation.
In summary the analysis of Fig.~(\ref{fig:supersymmetry}) shows that CEM is operative and $f_e / M_{Pl} \gg 1$ is achieved while $f < M_{Pl}$ consistent with WGC.

\section{Supergravity \label{sec:Supergravity}}
\begin{figure}
  \centering
  \begin{subfigure}{0.5 \textwidth}
    \includegraphics[width = \textwidth]{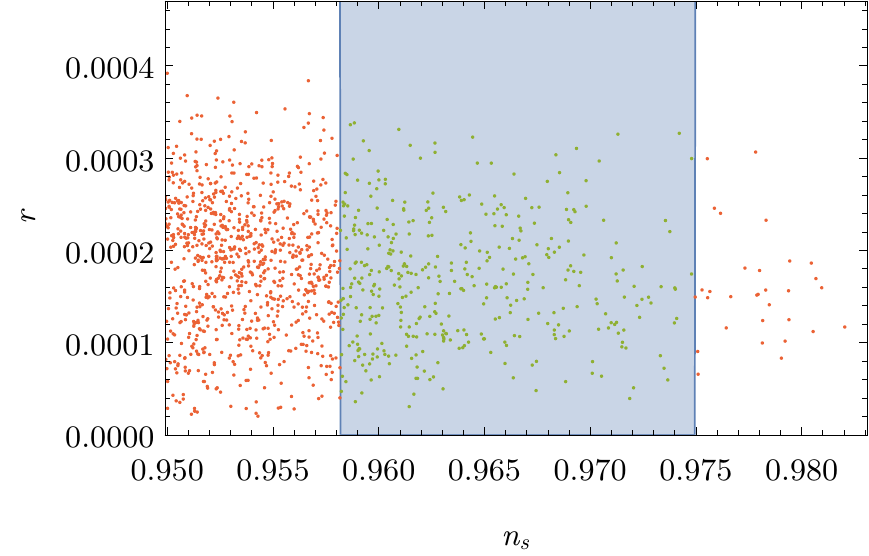}
  \end{subfigure}
  \begin{subfigure}{0.5 \textwidth}
    \includegraphics[width = \textwidth]{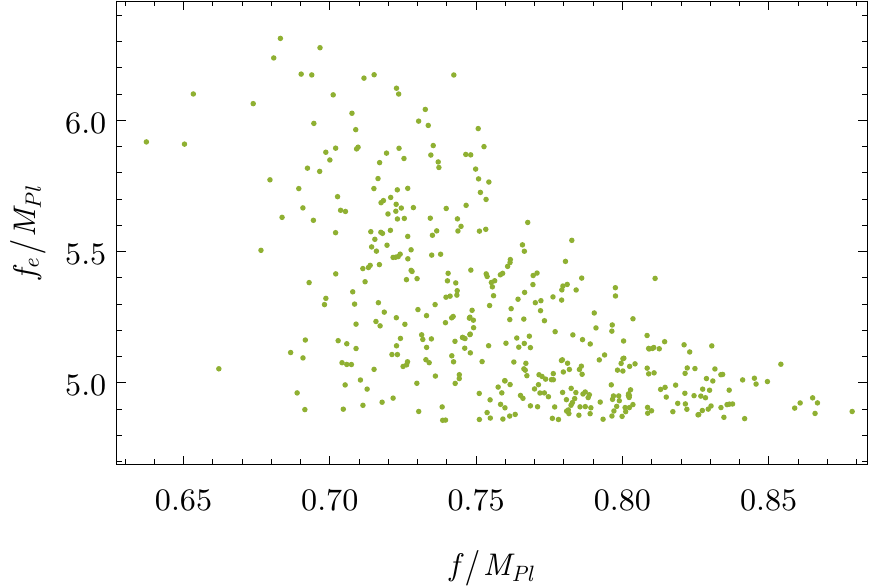}
  \end{subfigure}
  \begin{subfigure}{0.55 \textwidth}
    \includegraphics[width = \textwidth]{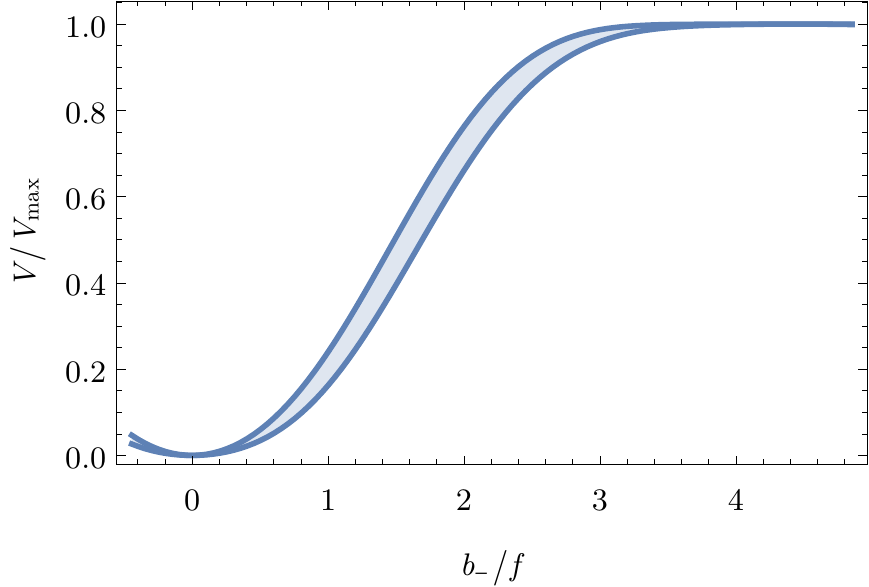}
  \end{subfigure}
  \caption{\protect\input{figures/supergravity.txt}
    Top panel: Plot of $r$ vs $n_s$.
    The blue region contains parameter points that lie in the experimentally allowed range of $r$ and $n_s$.
    Middle panel: A plot of the effective axion decay constant $f_e / M_{Pl}$ vs $f / M_{Pl}$ for the parameter points that lie in the blue region in the left panel.
    Bottom panel: Plot of $V / V_\text{max}$ vs $b_- / f$ for the parameter points that lie in the blue region in the top panel.
    As in the global supersymmetry analysis of Fig.~(\ref{fig:supersymmetry}) the supergravity analysis here shows that axion inflation for the set of points in the blue region is consistent with WGC as exhibited by the middle panel.
  } \label{fig:supergravity}
\end{figure}

Next we test CEM for supergravity where the scalar potential has the form~\cite{Chamseddine:1982jx, Cremmer:1982en}
\begin{equation} \label{eq:supergravity:potential}
  V = e^{K / M_{Pl}^2} \left[
    D_i W K^{-1}_{ij^*} D_{j^*} W^* - \frac{3}{M_{Pl}^2} \left|W\right|^2
  \right] + V_D\,,
\end{equation}
where $K$ is the K\"ahler potential, $W$ as before is the superpotential, and $D_i W$ is defined by
\begin{equation} \label{eq:supergravity:DW}
  D_i W = \frac{\partial W}{\partial \phi_i}
        + \frac{1}{M_{Pl}^2} \frac{\partial K}{\partial \phi_i} W\,.
\end{equation}
$V_D$, which is the $D$-term of the potential, will play no role in our analysis and will be dropped from here on.
In order to avoid the so-called $\eta$-problem of supergravity we choose the K\"ahler potential to be of the form
\begin{equation}
  K = \sum_i \frac{1}{2} \left(\Phi_i + \Phi_i^\dagger\right)^2\,,
\end{equation}
where we consider a pair of chiral fields $\Phi_i, i = 1, 2$.
We parametrize the complex scalar components $\phi_i$ of the fields as
\begin{equation}
  \phi_i = \left(\rho_i + i a_i\right) / \sqrt 2,
  ~~~ i = 1, 2\,,
\end{equation}
where $a_i$ have the shift symmetry
\begin{equation}
  a_1 \to a_1 + \lambda,
  ~~~ a_2 \to a_2 - \lambda\,,
\end{equation}
and $\rho_i$ are the saxion fields.
It is then easily checked that the kinetic energy for $\phi_i$ and $a_i$ is canonically normalized.
As in the global supersymmetry case we choose $W$ of the form Eq.~(\ref{eq:supersymmetry:W}) where, however, we write
\begin{equation}
  W_s = W_s^\text{vis} + W_0\,,
\end{equation}
where $W_s$ is invariant under the shift symmetry with $W_s^\text{vis}$ arising from the visible sector
\begin{equation}
  W_s^\text{vis} =
      \frac{\mu}{2} \left(\Phi_1 + \Phi_2\right)^2
    + \frac{\lambda}{3} \left(\Phi_1 + \Phi_2\right)^3\,,
\end{equation}
and $W_0$ arising from the hidden sector. We set $W_0$ in a way that $W = 0$ if $a_i = 0$, which ensures vanishing of the vacuum energy at the end of inflation.
For supergravity analysis the saxion can be stabilized by imposition of spontaneous symmetry breaking conditions~\cite{Nath:1983aw}
\begin{equation}
  D_i W = 0,
  ~~~ i = 1, 2\,.
\end{equation}
For shift symmetry breaking we assume
\begin{equation}
  W_{sb} = \sum_{n = 1}^q A_n \left(e^{c_n \Phi_1} + e^{c_n \Phi_2}\right)\,,
\end{equation}
and as in the global supersymmetry case we make a change of basis from $a_1, a_2$ to $b_+, b_-$ as given by Eq.~(\ref{eq:b+-}), where $a$ and $\bar a$ are replaced with $a_1$ and $a_2$ respectively.
Next we expand around the minimum of the saxion potential and retain only $b_-$ which controls the slow-roll part of the potential.
To that end $W_{sb}$ takes the form
\begin{equation}
  W_{sb} = \sum_{n = 1}^q A_n \left(
      e^{i \gamma_n \frac{b_-}{\sqrt{2} f}}
    + e^{-i \gamma_n \frac{b_-}{\sqrt{2} f}}
  \right)\,,
\end{equation}
where we take $\gamma_n = c_n f / \sqrt{2}$.
In this case the slow-roll part of the potential which involves only the field $b_-$ takes the form~\cite{Nath:2017ihp}
\begin{equation} \label{eq:supergravity:Vslow}
  \begin{aligned}
    V\left(b_-\right) =
      & 4 M_{Pl}^4 e^{2 f^2 / M_{Pl}^2} \sum_{n = 1}^q \sum_{m = 1}^q
        e^{\gamma_n + \gamma_m} \frac{A_n A_m}{M_{Pl}^6}\\
        &{} \times \left[
          \gamma_n \gamma_m \frac{M_{Pl}^2}{f^2} \left(
              1
            - \cos\left(\gamma_n \frac{b_-}{\sqrt{2} f}\right)
            - \cos\left(\gamma_m \frac{b_-}{\sqrt{2} f}\right)
            + \cos\left(\left(\gamma_n - \gamma_m\right) \frac{b_-}{\sqrt{2} f}\right)
          \right)\right.\\
          &~~~~~~ + \left(2 \gamma_n + 2 \gamma_m - 3 + 4 \frac{f^2}{M_{Pl}^2}\right) \left(
              1
            - \cos\left(\gamma_n \frac{b_-}{\sqrt{2} f}\right)
            - \cos\left(\gamma_m \frac{b_-}{\sqrt{2} f}\right)\right.\\
            &~~~~~~~~~~~~ \left.\left.{}
            + \frac{1}{2} \cos\left(\left(\gamma_n - \gamma_m\right) \frac{b_-}{\sqrt{2} f}\right)
            + \frac{1}{2} \cos\left(\left(\gamma_n + \gamma_m\right) \frac{b_-}{\sqrt{2} f}\right)
          \right)
        \right]\,.
  \end{aligned}
\end{equation}
For our analysis we take $\gamma_n = n$ and $q = 3$, which is the minimal value with which we were able to achieve experimentally-consistent inflation.
In that case the above potential consists of a superposition of six cosines so that
\begin{equation} \label{eq:supergravity:Vslow3}
  V\left(b_-\right)
    = M_{Pl}^4 e^{2 f^2 / M_{Pl}^2} \sum_{k = 1}^6 C_k \left(1 - \cos\left(\frac{k b_-}{\sqrt{2} f}\right)\right)\,,
\end{equation}
where $C_k$ are given by
\begin{equation} \label{eq:supergravity:Vslow3Coefficients}
  \begin{aligned}
    C_1 &=   4 \left(
        2 e^2 \left(   \frac{M_{Pl}^2}{f^2} + 1 + 4 \frac{f^2}{M_{Pl}^2}\right) \frac{A_1^2  }{M_{Pl}^6}
      +   e^3 \left(                          3 + 4 \frac{f^2}{M_{Pl}^2}\right) \frac{A_1 A_2}{M_{Pl}^6}\right.\\
      &~~~~~~ \left.{}
      + 2 e^4 \left( 3 \frac{M_{Pl}^2}{f^2} + 5 + 4 \frac{f^2}{M_{Pl}^2}\right) \frac{A_1 A_3}{M_{Pl}^6}
      -   e^5 \left(12 \frac{M_{Pl}^2}{f^2} + 7 + 4 \frac{f^2}{M_{Pl}^2}\right) \frac{A_2 A_3}{M_{Pl}^6}
    \right)\,,\\
    C_2 &=   2 \left(
      -   e^2 \left(                          1 + 4 \frac{f^2}{M_{Pl}^2}\right) \frac{A_1^2  }{M_{Pl}^6}
      + 4 e^3 \left( 2 \frac{M_{Pl}^2}{f^2} + 3 + 4 \frac{f^2}{M_{Pl}^2}\right) \frac{A_1 A_2}{M_{Pl}^6}\right.\\
      &~~~~~~ \left.{}
      + 4 e^4 \left( 4 \frac{M_{Pl}^2}{f^2} + 5 + 4 \frac{f^2}{M_{Pl}^2}\right) \frac{A_2^2  }{M_{Pl}^6}
      - 2 e^4 \left( 6 \frac{M_{Pl}^2}{f^2} + 5 + 4 \frac{f^2}{M_{Pl}^2}\right) \frac{A_1 A_3}{M_{Pl}^6}\right.\\
      &~~~~~~ \left.{}
      + 4 e^5 \left( 6 \frac{M_{Pl}^2}{f^2} + 7 + 4 \frac{f^2}{M_{Pl}^2}\right) \frac{A_2 A_3}{M_{Pl}^6}
    \right)\,,\\
    C_3 &=   4 \left(
      -   e^3 \left(                          3 + 4 \frac{f^2}{M_{Pl}^2}\right) \frac{A_1 A_2}{M_{Pl}^6}
      + 2 e^4 \left( 3 \frac{M_{Pl}^2}{f^2} + 5 + 4 \frac{f^2}{M_{Pl}^2}\right) \frac{A_1 A_3}{M_{Pl}^6}\right.\\
      &~~~~~~ \left.{}
      + 2 e^5 \left( 6 \frac{M_{Pl}^2}{f^2} + 7 + 4 \frac{f^2}{M_{Pl}^2}\right) \frac{A_2 A_3}{M_{Pl}^6}
      + 2 e^6 \left( 9 \frac{M_{Pl}^2}{f^2} + 9 + 4 \frac{f^2}{M_{Pl}^2}\right) \frac{A_3^2  }{M_{Pl}^6}
    \right)\,,\\
    C_4 &=   2 \left(
      - 2 e^4 \left(                          5 + 4 \frac{f^2}{M_{Pl}^2}\right) \frac{A_1 A_3}{M_{Pl}^6}
      -   e^4 \left(                          5 + 4 \frac{f^2}{M_{Pl}^2}\right) \frac{A_2^2  }{M_{Pl}^6}
    \right)\,,\\
    C_5 &= - 4
          e^5 \left(                          7 + 4 \frac{f^2}{M_{Pl}^2}\right) \frac{A_2 A_3}{M_{Pl}^6}\,,\\
    C_6 &= - 2
          e^6 \left(                          9 + 4 \frac{f^2}{M_{Pl}^2}\right) \frac{A_3^2  }{M_{Pl}^6}\,.
  \end{aligned}
\end{equation}
As in the global supersymmetry case of section \ref{sec:Supersymmetry} here also local regions of flatness in the potential appear due to overlaps of several cosines, slow roll can occur and CEM is operative.
Simulation for this potential is similar to the global supersymmetry model and are shown on Fig.~(\ref{fig:supergravity}).
Further one finds that there exist regions of the parameter space where $f_e \gg M_{Pl}$ while $f < M_{Pl}$ consistent with WGC.
In this model, we find $\left|f_e - f_{eH}\right| / f_e \le \protect 2\%$.

\section{Dirac-Born-Infeld (DBI) \label{sec:DBI}}
In sections \ref{sec:Supersymmetry} and \ref{sec:Supergravity} we discussed applications of CEM which accomplishes successful axion inflation in conformity with the current cosmological data and consistent with WGC.
As mentioned in section \ref{sec:Introduction} CEM works only for models where slow roll is governed by flat potentials.
This, however, is not the case for DBI-flation.
Here the entire Lagrangian enters in slow roll and CEM is not applicable.
Nonetheless we will show in this section that axionic DBI-flation allows for successful inflation consistent with cosmological data and consistent with WGC.
We will work within the framework of supersymmetric DBI actions which have been investigated by a number of authors (see, e.g.,~\cite{Nath:2018xxe, Khoury:2010gb, Khoury:2011da, Baumann:2011nk, Baumann:2011nm, Rocek:1997hi, Tseytlin:1999dj, Ito:2007hy, Billo:2008sp, Sasaki:2012ka, Aoki:2016tod}.
Inflation in a single field DBI was discussed in~\cite{Sasaki:2012ka} and for the case of two fields in~\cite{Nath:2018xxe}.
Thus as in our analysis in sections 4 and 5 we consider a pair of chiral superfields $\Phi_1$ and $\Phi_2$ which carry opposite charges under a global $U\left(1\right)$ symmetry.
The supersymmetric Lagrangian involving $\Phi_1$ and $\Phi_2$ is given by
\begin{equation} \label{eq:DBI:lagrangianTerms}
  \mathcal{L} = \mathcal{L}_D + \mathcal{L}_F\,,
\end{equation}
where $\mathcal{L}_D$ is the $D$-part of the Lagrangian and $\mathcal{L}_F$ is the $F$-part.
Here $\mathcal{L}_D$ consists of a part which is quadratic in the fields and a part which is quartic in the fields so that
\begin{equation} \label{eq:DBI:lagrangianD}
  \mathcal{L}_D = \sum_{k = 1}^2 \left(\int d^4 \theta \Phi_k \Phi_k^\dagger
    + \int d^4 \theta \frac{\alpha_1}{16}
      \left(D^\alpha \Phi_k D_\alpha \Phi_k\right)
      \left({\bar D}^{\dot\alpha} \Phi_k^\dagger {\bar D}_{\dot\alpha} \Phi_k^\dagger\right)
      G\left(\phi\right)\right)\,,
\end{equation}
where
\begin{equation}
  G\left(\phi\right) = \frac{1}{T} \frac{1}{1 + P + \sqrt{\left(1 + P\right)^2 - Q}}\,,
\end{equation}
and $P$ and $Q$ are assumed to have the following forms
\begin{equation} \label{eq:DBI:PQ}
  \begin{aligned}
    P &= \left(
        \partial_a \phi_1 \partial^a \phi^*_1
      + \partial_a \phi_2 \partial^a \phi^*_2
    \right) / T\,,\\
    Q &= \left(
        \alpha_1 \partial_a \phi_1 \partial^a \phi_1 \partial_b \phi^*_1 \partial^b \phi^*_1
      + \alpha_1 \partial_a \phi_2 \partial^a \phi_2 \partial_b \phi^*_2 \partial^b \phi^*_2
    \right) / T^2\,.
  \end{aligned}
\end{equation}
Here $T$ is a parameter of dimension 4 in mass and can be thought of as a warp factor arising from a higher dimensional geometry.
We note that the Lagrangian of Eq.~(\ref{eq:DBI:lagrangianD}) is a direct generalization of the Lagrangian for the single field case which can be derived from a more basic 3-brane action (see, e.g.,~\cite{Rocek:1997hi, Tseytlin:1999dj, Sasaki:2012ka} and the references therein).
In writing Eq.~(\ref{eq:DBI:lagrangianD}) we imposed an additional constraint which is invariance under $\Phi_1$ and $\Phi_2$ interchange.
Finally $\mathcal{L}_F$ is given by
\begin{equation}
  \mathcal{L}_F = \int d^2 \theta W\left(\Phi_1, \Phi_2\right)
                + \int d^2 \bar\theta W^*\left(\Phi_1^\dagger, \Phi_2^\dagger\right)\,,
\end{equation}
where the superpotential $W$ as in earlier analyses is given by $W = W_s + W_{sb}$, and where
\begin{equation}
W_s = \mu \Phi_1 \Phi_2 + \frac{\lambda}{2 M_{Pl}} \left(\Phi_1 \Phi_2\right)^2
\end{equation}
is chosen so that we can stabilize the saxion VEVs and $W_{sb}$ breaks the global $U\left(1\right)$ symmetry and is taken to be of the form
\begin{equation} \label{eq:dbi:Wsb}
  W_{sb} = \sum_{k = 1}^m \left(A_{k} \Phi_1^k + A_{k} \Phi_2^k\right)\,.
\end{equation}
Integration over the Grassmann variables gives rise to the following Lagrangian
\begin{equation} \label{eq:dbi:lagrangianIntermediate}
  \begin{aligned}
    \mathcal{L} =
      & T - T \sqrt{\left(1 + P\right)^2 - Q} + F_1 F^*_1 + F_2 F^*_2\\
      &+ G\left(\phi\right) \left[
        \alpha_1 \left(
          - 2 F_1 F^*_1 \partial_a \phi_1 \partial^a \phi^*_1
          + F_1^2 {F^*_1}^2
        \right)\right.\\
        &~~~ \left.{} + \alpha_1 \left(
          - 2 F_2 F^*_2 \partial_a \phi_2 \partial^a \phi^*_2
          + F_2^2 {F^*_2}^2
        \right)\right]\\
      &+ \left(
          \frac{\partial W}{\partial \phi_1} F_1
        + \frac{\partial W}{\partial \phi_2} F_2
        + h.c.
      \right)\,.
  \end{aligned}
\end{equation}
There are four auxiliary fields in Eq.~(\ref{eq:dbi:lagrangianIntermediate}) which are $F_1$, $F^*_1$, $F_2$, $F^*_2$.
The auxiliary fields $F_k$ satisfy the cubic equation
\begin{equation}
  F_k^3 + p_k F_k + q_k = 0\,,
  ~~~ k = 1, 2\,,
\end{equation}
where $p_k$, $q_k$ are defined by
\begin{equation} \label{eq:DBI:pq}
  \begin{aligned}
    p_k &=
      \left(\frac{\partial W}{\partial \phi_k}\right)^{-1}
      \frac{\partial W^*}{\partial \phi^*_k}
      \frac
        {1 - 2 \alpha_1 G\left(\phi\right) \partial_\mu \phi_k \partial^\mu \phi_k}
        {2 \alpha_1 G\left(\phi\right)}\,,\\
    q_k &=
      \frac{1}{2 \alpha_1 G\left(\phi\right)}
      \left(\frac{\partial W}{\partial \phi_k}\right)^{-1}
      \left(\frac{\partial W^*}{\partial \phi^*_k}\right)^2\,.
  \end{aligned}
\end{equation}
Since $F_k$ satisfies a cubic equation, it has three roots which are
\begin{equation} \label{eq:DBI:F}
  \begin{aligned}
    F_k = &\omega^j \left(
      - \frac{q_k}{2}
      + \sqrt{\left(\frac{q_k}{2}\right)^2 + \left(\frac{p_k}{3}\right)^3}\right)^{1 / 3}\\
    & + \omega^{3 - j} \left(
      - \frac{q_k}{2}
      - \sqrt{\left(\frac{q_k}{2}\right)^2 + \left(\frac{p_k}{3}\right)^3}\right)^{1 / 3}\,,
  \end{aligned}
\end{equation}
where $\omega$ is the cube root of unity and $j = 0, 1, 2$.
It turns out that of the three roots only $j = 0$ is a solution to the full Euler-Lagrange equations for $F_k$ and in our analysis we consider only this solution.

An explicit computation of the Lagrangian in this case is given in~\cite{Nath:2018xxe} and displayed in Eq.~(\ref{eq:dbi:lagrangian}).
The Lagrangian depends on a single axion field $b_-$ defined as in the preceding sections and 5 parameters $T$, $\alpha_1$, $f$, $\tilde\beta$, and a vector $\mathcal{G}$ as discussed below.
Thus we have
\begin{equation} \label{eq:dbi:lagrangian}
  \begin{aligned}
    &\mathcal{L}\left(T, \alpha_1, f, \beta, G; b_-, \dot{b_-}\right) = T \left(
        1
      - \sqrt{
          1
        - \frac{{\dot b}_-^2}{T}
        + \frac{\left(2 - \alpha_1\right) {\dot b}_-^4}{8 T^2}
      }\right.\\
      &~~~ \left.{}
      + 2 \mathcal{F}_+^2
      + 2 \mathcal{F}_-^2
      - \frac{4}{3 \alpha_1} \left(
        \mathcal{T} + \left(\alpha_1 - 1\right) \frac{{\dot b}_-^2}{4 T}
      \right)
      + 4 k \left(\mathcal{F}_+ + \mathcal{F}_-\right)\right.\\
      &~~~ \left.{}
      + \frac{\alpha_1}{\mathcal{T} - {\dot b}_-^2 / \left(4 T\right)}\left(
          2 \left(
              \mathcal{F}_+^2
            + \mathcal{F}_-^2
            - \frac{2}{3 \alpha_1}
              \left(\mathcal{T} + \left(\alpha_1 - 1\right) \frac{{\dot b}_-^2}{4 T}\right)
          \right) \frac{{\dot b}_-^2}{4 T}
        + \mathcal{F}_+^4\right.\right.\\
        &~~~~~~ \left.\left.{}
        + \mathcal{F}_-^4
        + \frac{2}{3 \alpha_1^2}
          \left(\mathcal{T} + \left(\alpha_1 - 1\right) \frac{{\dot b}_-^2}{4 T}\right)^2
        - \frac{4}{3 \alpha_1}
          \left(\mathcal{T} + \left(\alpha_1 - 1\right) \frac{{\dot b}_-^2}{4 T}\right)
          \left(\mathcal{F}_+^2 + \mathcal{F}_-^2\right)
      \right)
    \right)\,,
  \end{aligned}
\end{equation}
where
\begin{equation}
  \begin{aligned}
    \mathcal{F}_\pm = \pm &\left(
      \mp\frac{1}{2 \alpha_1} k \left(\mathcal{T} - \frac{{\dot b}_-^2}{4 T}\right)\right.\\
      &\left.{} + \sqrt{
          \frac{1}{4 \alpha_1^2} k^2 \left(\mathcal{T} - \frac{{\dot b}_-^2}{4 T}\right)^2
        + \frac{1}{27 \alpha_1^3} \left(
            \mathcal{T}
          + \left(\alpha_1 - 1\right) \frac{{\dot b}_-^2}{4 T}
        \right)^3
      }
    \right)^{1 / 3}\,,
  \end{aligned}
\end{equation}
and where
\begin{align} 
  \mathcal{T} &= \frac{1}{2} \left(
      1
    + \sqrt{1 - \frac{{\dot b}_-^2}{T} + \frac{\left(2 - \alpha_1\right){\dot b}_-^4}{8 T^2}}
  \right)\,,\\
  \label{eq:dbi:beta}
  k &= \tilde\beta \sqrt{\sum_{m, n} m n \mathcal{G}_m \mathcal{G}_n \left(
      1
    - \cos\left(\frac{b_- m}{\sqrt{2} f}\right)
    - \cos\left(\frac{b_- n}{\sqrt{2} f}\right)
    + \cos\left(\frac{b_- \left(m - n\right)}{\sqrt{2} f}\right)
  \right)}\,,\\
  \mathcal{G}_k &= \frac{A_k 2^{1 / 2 \left(1 - k\right)}}{\tilde\beta \sqrt{T} f^{1 - k}}\,.
\end{align}

We note that the parameter $\tilde\beta$ here is redundant, and is chosen in such a way as to make $\mathcal{G}_k \sim 1$.
The first non-zero component of $\mathcal{G}$ can also be set to $1$ to reduce redundancy.
Further, we note that Eqs.~(\ref{eq:epsEtaFromPotential}) will not be sufficient to describe slow roll in this case, because they do not take the form of kinetic energy into account.
However, we will use Eqs.~(\ref{eq:slowRollParametersDynamic}) which are more general.
Thus we conjecture that while Eqs.~(\ref{eq:slowRollParametersDynamicFromStatic}) do not hold, Eq.~(\ref{eq:feFromDynamicSlowRollParameters}) can still be used to derive an effective decay constant, where Eqs.~(\ref{eq:slowRollParametersDynamic}) are used to derive $\epsilon_H$ and $\eta_H$.

\begin{figure}
  \centering
  \begin{subfigure}{0.5 \textwidth}
    \includegraphics[width = \textwidth]{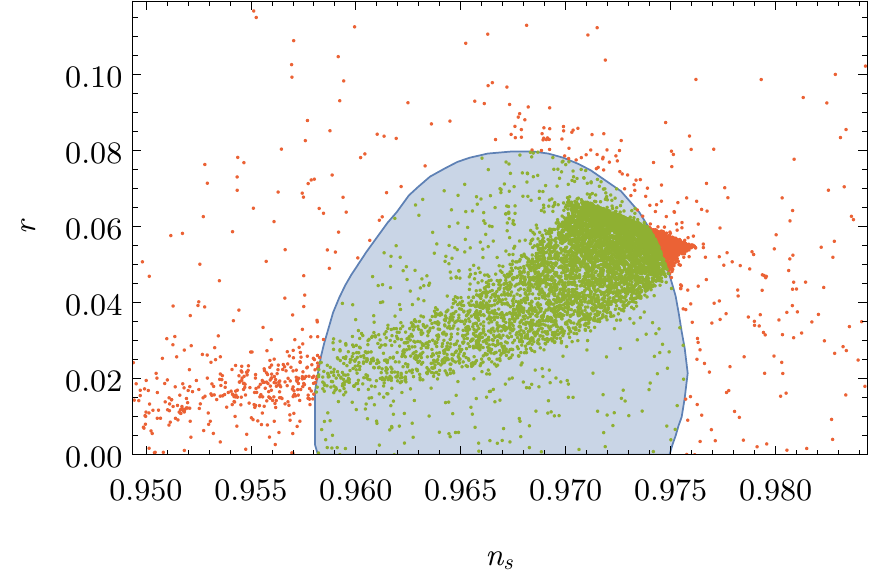}
  \end{subfigure}
  \begin{subfigure}{0.5 \textwidth}
    \includegraphics[width = \textwidth]{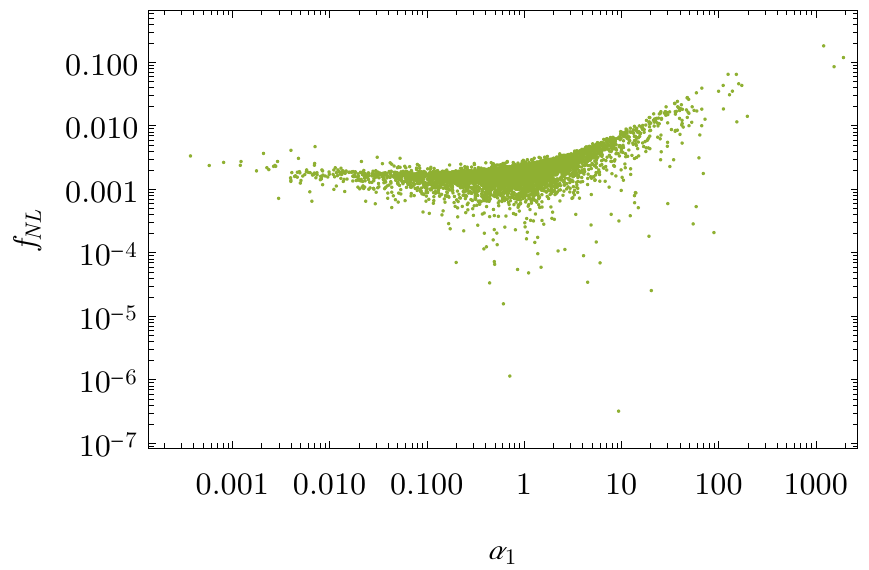}
  \end{subfigure}
  \begin{subfigure}{0.5 \textwidth}
    \includegraphics[width = \textwidth]{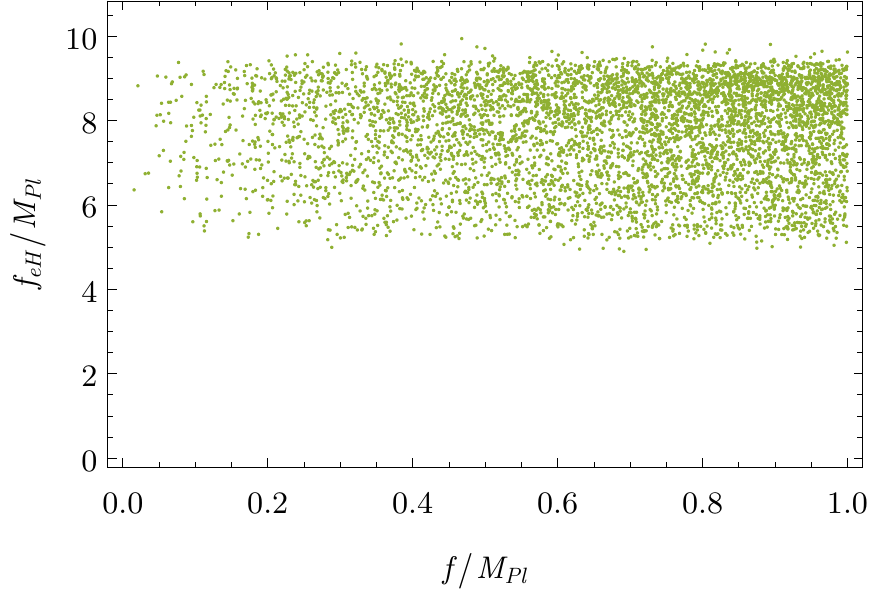}
  \end{subfigure}
  \caption{\protect\input{figures/DBI.txt}
    Top panel is a plot of $r$ vs $n_s$.
    Middle panel shows the non-Gaussianity parameter $f_{NL}^\text{equil}$ as a function of $\alpha_1$ (for a discussion of $f_{NL}^\text{equil}$ see ~\cite{Nath:2018xxe}).
    Bottom panel is a plot of $f_{eH}$ vs $f$ where $f_{eH}$ is defined by Eq.~(\ref{eq:feFromDynamicSlowRollParameters}).
    We note that all the points in the blue of the top panel have $f < M_{Pl}$ as exhibited in the bottom panel which ensures that DBI-flation is consistent with WGC.
  } \label{fig:DBI}
\end{figure}

Simulation for DBI is shown on Fig.~(\ref{fig:DBI}).
We sample the parameter space of the model Eq.~(\ref{eq:dbi:lagrangian}) by setting $T = 10^{-12} M_{Pl}^4$, $\mathcal{G}_1 = \mathcal{G}_2 = \mathcal{G}_3 = 0$, $\mathcal{G}_4 = 1$, and varying $\mathcal{G}_5$, $\mathcal{G}_6$, $\alpha_1$, $f$, $\tilde\beta$, and the pivot e-foldings count $N_\text{pivot}$.
Here we use the data of Fig~(1) of~\cite{Nath:2018xxe}, but with an update of the experimental constraints on $r$ and $n_s$ as given by Planck 2018 results~\cite{Akrami:2018odb}, and compute $f_{eH}$ using Eq.~(\ref{eq:feSpectralIndices}).
Note that even though $f_e$ cannot be defined in this model, $f_{eH}$ still satisfies Eq.~(\ref{eq:feExperimentalConstraint}).
The top panel of Fig.~(\ref{fig:DBI}) shows the parameter points of the DBI model which lie in the experimentally allowed domain (the blue region).
The middle panel gives $f^{\text{equil}}_{\text{NL}}$ as function of $\alpha_1$.
Recently, the Planck Collaboration~\cite{Akrami:2019izv} has analyzed the Planck full-mission cosmic microwave background (CMB) temperature and E-mode polarization maps to obtain constraints on primordial non-Gaussianity.
Their combined temperature and polarization analysis produces the following final result on $f^{\text{equil}}_{\text{NL}}$ so that $f^{\text{equal}}_{\text{NL}} = -26 \pm 47 \left(68\%\, \text{CL, statistical}\right)$.
We note that while the prediction of $f^{\text{equil}}_{\text{NL}}$ as given by the middle panel of Fig.~(\ref{fig:DBI}) is consistent with experiment, it is far too small to be tested in the near future.
The bottom panel of Fig.~(\ref{fig:DBI}) gives a plot of $f_{eH} / M_{Pl}$ vs $f / M_{Pl}$.
One finds that $f_{eH} / M_{Pl} \gg 1$ while $f / M_{Pl} < 1$ consistent with WGC.

\section{Slow-roll parameters in terms of density and the size of $r$ in single-field models \label{sec:r}}
It is interesting to observe that the tensor-to-scalar ratio $r$ in the effective single field models of global supersymmetry in Fig.~(\ref{fig:supersymmetry}) and of supergravity in Fig.~(\ref{fig:supergravity}) is $O\left(10^{-4}\right)$, while for the DBI case Fig.~(\ref{fig:DBI}) it is much larger than that and for some parameter points it can be as large as the current experimental upper limit $r = 0.07$.
To see how this is possible, we consider first an arbitrary single-field inflation model with canonical kinetic energy.
Here for the number of e-foldings we have
\begin{equation} \label{eq:efoldingsGeneral}
  N = \frac{1}{M_{Pl}} \int \sqrt{\rho / 3}\,dt
    = \frac{1}{M_{Pl}} \int \sqrt{\rho / 3}\,\frac{d\phi}{\dot \phi}\,,
\end{equation}
where $\rho$ is the density of the inflaton field $\phi$.
In slow-roll approximation $\ddot \phi \approx 0$, and ${\dot \phi}^2 \ll V\left(\phi\right)$, and using equations of motion we have
\begin{equation} \label{eq:efoldingsCanonical}
  N = - \frac{1}{M_{Pl}^2} \int \frac{V\left(\phi\right)}{V^\prime\left(\phi\right)} d\phi
    \approx - \frac{\Delta \phi}{M_{Pl} \sqrt{2 \epsilon}}\,,
\end{equation}
where we used the definition of the slow-roll parameter $\epsilon$, see Eq.~(\ref{eq:epsEtaFromPotential}).
This leads to a relation between $r$ and $\Delta \phi$ so that
\begin{equation} \label{eq:Deltaphi}
  r \approx \frac{8 \Delta\phi^2}{M_{Pl}^2 N^2}\,.
\end{equation}

\begin{figure}
  \centering
  \begin{subfigure}{0.5 \textwidth}
    \includegraphics[width = \textwidth]{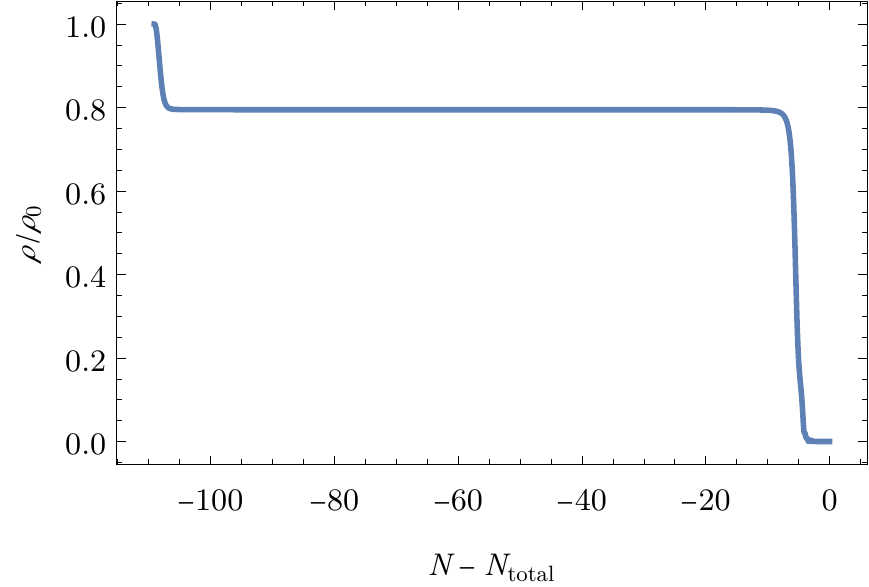}
  \end{subfigure}
  \begin{subfigure}{0.5 \textwidth}
    \includegraphics[width = \textwidth]{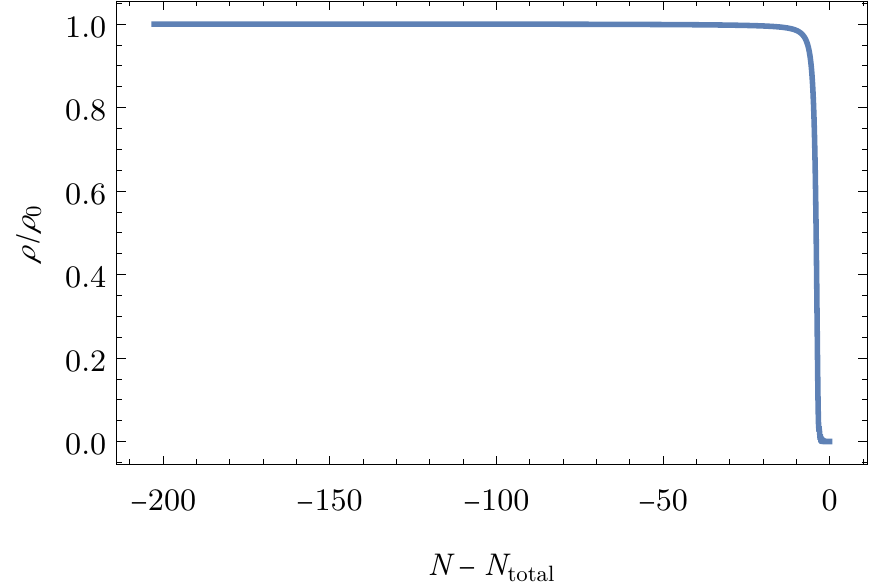}
  \end{subfigure}
  \begin{subfigure}{0.5 \textwidth}
    \includegraphics[width = \textwidth]{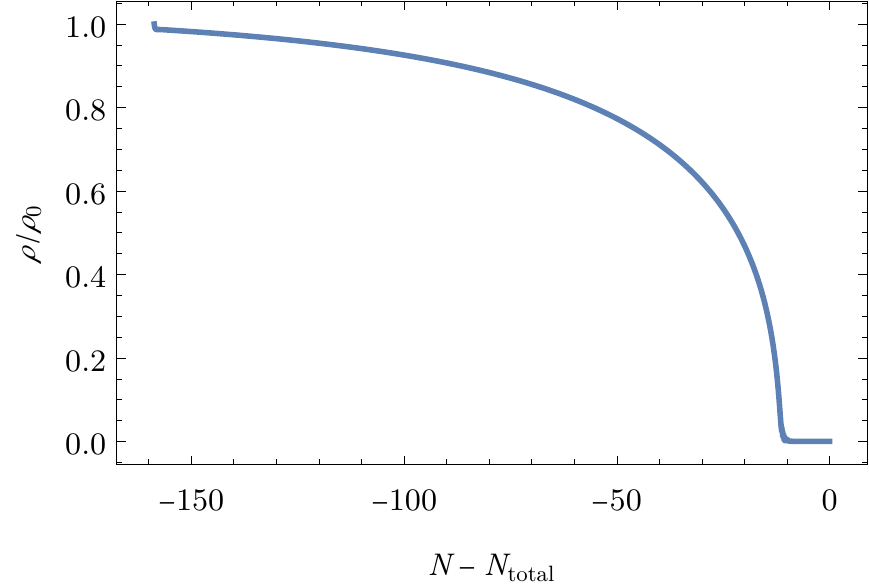}
  \end{subfigure}
  \caption{
    Top panel: Evolution of $\rho / \rho_0$, where $\rho_0$ is the density at the beginning of simulation, as a function of the number of e-foldings until the end of inflation for global supersymmetry case.
    For convenience the number of e-foldings is shown negative as the x-axis records $N - N_\text{total} < 0$.
    Horizon exit occurs at $-60 < N - N_\text{total} < -50$.
    The plot is for one of the global supersymmetry parameter sets.
    Middle panel: Same as the left panel except it is for the supergravity case.
    Bottom panel: The same as the top panel except it is for the DBI case.
    The figure explains the reason why $r$ can be much larger for DBI relative to models where slow roll is controlled by flat potentials.
    From Eq.~(\ref{eq:rFromDDensityDN}) we see that $r$ is proportional to $d\rho / dN$.
    In the top and middle panels $\rho / \rho_0$ is essentially flat near the horizon exist which is in the interval which occurs at $-60 < N - N_\text{total} < -50$ and leads to $r \sim O\left(10^{-4}\right)$ while for the DBI case in the bottom panel $\rho / \rho_0$ has a significant curvature and $r$ can be as large as the experimental upper limit of $0.07$.
  } \label{fig:density}
\end{figure}
We consider now the global supersymmetry case where $\Delta\phi / f < 1$ and $f / M_{Pl} < 1$ (for consistency with WGC, see Figs.~(\ref{fig:supersymmetry})) and $N = \left[50 - 60\right]$ which implies
\begin{equation} \label{eq:rBound}
  r < 0.003\,.
\end{equation}
A similar analysis holds for the supergravity case.
The analyses on $r$ in Figs.~(\ref{fig:supersymmetry}, \ref{fig:supergravity}) show that $r \sim O\left(10^{-4}\right)$ is consistent with the bound of Eq.~(\ref{eq:rBound}).
The reason for the smallness of $r$ in Figs.~(\ref{fig:supersymmetry}, \ref{fig:supergravity}) and more generally for models with flat potentials can be traced to Eq.~(\ref{eq:Deltaphi}) and the condition $\Delta\phi < M_{Pl}$.

However, Eq.~(\ref{eq:Deltaphi}) is not applicable to DBI-flation and k-flation~\cite{Garriga:1999vw}.
Here we need to look at the evolution of the density $\rho$ of the inflaton field as a function of e-foldings.
Thus for the case when sound speed $c_s \sim 1$, one can obtain the slow-roll parameters in terms of density and its derivatives with respect to the number of e-foldings by using the Friedmann equation
\begin{equation}
  H = \dot N = \frac{1}{M_{Pl}} \sqrt{\frac{\rho}{3}}\,,
\end{equation}
and
\begin{equation}
  \dot\rho = \frac{d\rho}{dN} \dot{N}\,.
\end{equation}
Using the above together with Eq.~(\ref{eq:slowRollParametersDynamic}) we get
\begin{equation}
  \begin{aligned}
    \epsilon_H &= -\frac{1}{2 \rho} \frac{d\rho}{dN}\,,\\
    \eta_H &= -\frac{1}{\rho} \frac{d\rho}{dN} + \left.\frac{d^2\rho}{dN^2} \middle/ \frac{d\rho}{dN}\right.\,.
  \end{aligned}
\end{equation}
Further, the enhancement factor $f_{eH}$ is given by
\begin{equation}
  f_{eH} = M_{Pl} \sqrt{\left.\frac{d\rho}{dN} \middle/ \frac{d^2\rho}{dN^2}\right.}\,.
\end{equation}
Similarly the spectral index $n_s$ and the ratio $r$ of the tensor to scalar power spectrum are given by
\begin{equation} \label{eq:rFromDDensityDN}
  \begin{aligned}
    n_s &= 1 + \frac{2}{\rho} \frac{d\rho}{dN} - \left.\frac{d^2\rho}{dN^2} \middle/ \frac{d\rho}{dN}\right.\,,\\
    r &= - \frac{8}{\rho} \frac{d\rho}{dN}\,.
  \end{aligned}
\end{equation}

From the plot of $\rho / \rho_0$ as a function of $N - N_{\text{total}}$ one finds that $\frac{d\rho}{dN}$ is very small for the top and middle panels of Fig.~(\ref{fig:density}) in the domain of horizon exit, i.e., $-60 < N - N_\text{total} < -50$ and leads to $r \sim O\left(10^{-4}\right)$.
On the other hand for the DBI case $\frac{d\rho}{dN}$ is visibly much larger as can be seen by the bottom panel of Fig.~(\ref{fig:density}).
This explains why $r$ is much larger for DBI-flation than for the case where inflation is driven by a flat potential which is the case for the top and the middle panels of Fig.~(\ref{fig:density}).

\section{Conclusion \label{sec:Conclusion}}
One of the possible candidates for an inflaton is an axion.
However, axion models with a simple cosine potential require an axion decay constant which is super-Planckian in size which is in conflict with the Weak Gravity Conjecture.
In this work we propose a new mechanism, the Coherent Enhancement Mechanism, which allows one to produce an effective decay constant which governs inflation to be much larger than the true decay constant that enters in the microscopic Lagrangian.
However, CEM works for the class of models where slow roll is governed by a flat potential.
To check the validity of CEM we work in a landscape of chiral superfields where the microscopic Lagrangian possesses a $U\left(1\right)$ global shift symmetry which is broken by instanton type terms.
The inflaton is identified with the pseuso-Nambu-Goldstone boson (pNGB) which is the lightest field in the broken $U\left(1\right)$ symmetry phase.
The proposed mechanism to enhance the effective axion decay constant associated with pNGB utilizes coherence among several cosines in the effective pNGB potential to produce a locally flat potential where slow roll can occur.
In this work we have illustrated CEM for models based in supersymmetry and in supergravity where inflation is driven by the pNGB potential.
In these analyses we show that successful inflation consistent with the cosmological data on the spectral indices and the tensor to scalar power spectrum ratio can be achieved along with consistency with WGC.
However, CEM is not valid for DBI-flation and more generally for k-flation.
Here inflation is governed not just by the potential but by the full Lagrangian.
Nonetheless our analysis shows that a successful DBI-flation can be achieved consistent with cosmological data and consistent with WGC.
It is also seen that while for single field inflation $r \sim O\left(10^{-4}\right)$, for DBI-flation one may have $r$ as large as the current experimental upper limit of $r = 0.07$.
An explanation of this phenomenon is given in section \ref{sec:r}.
We also show that the effective decay constant can be directly related to the spectral indices as exhibited by Eq.~(\ref{eq:feSpectralIndices}).
The analysis presented in this work shows that in all the cases considered axion inflation consistent with the experimental data can be accomplished with the axion decay constant in the microscopic Lagrangian in the sub-Planckian domain in conformity with the Weak Gravity Conjecture.\\~\\~\\

\textbf{Acknowledgments:}
Conversations with James Halverson and Cody Long are acknowledged.
This research was supported in part by the NSF Grant PHY-1620575.

\clearpage


\end{document}